\documentclass[letterpaper,12pt]{article}

\usepackage{mathptmx}

\usepackage{graphics}

\usepackage{geometry}
\usepackage{color}
\usepackage{amsmath}
\usepackage{amssymb}
\usepackage[markup=underlined]{changes}
\usepackage[authoryear,comma,longnamesfirst,sectionbib]{natbib} 

\usepackage{tikz}
\usepackage{amsthm}
\newtheorem{theorem}{Theorem}
\newtheorem{remark}{Remark}

\newtheorem*{proof*}{Proof}

\newtheorem{example}{Example}

\newtheorem{assumption}{Assumption}

\newcommand*{\indep}{%
\bot
}
\newcommand*{\nindep}{%
	\mathbin{%                   % The final symbol is a binary math operator
		\mathpalette{\@indep}{\not}% \mathpalette helps for the adaptation
	}%
}

\begin{document}
	\title{Estimating Average Treatment Effects Utilizing Fractional Imputation  when Confounders are Subject to Missingness}
	\author{Nathan Corder\thanks{North Carolina State University; necorder@ncsu.edu}  \&  Shu Yang\thanks{North Carolina State University; syang24@ncsu.edu}}
	\maketitle
	
	%%Missingness in observational data is ubiquitous. When the confounders are missing at random, multiple imputation is commonly used; however, the method requires congeniality conditions for valid inferences, which may not be satisfied when estimating average treatment effects. Alternatively, fractional imputation, proposed by Kim 2011, has been implemented to handling missing values in regression context. In this article, we develop fractional imputation methods for estimating the average treatment effects with confounders missing at random. We show that the fractional imputation estimator of the average treatment effect is asymptotically normal, which permits a consistent variance estimate. Via simulation study, we compare fractional imputation's accuracy and precision with that of multiple imputation. 
	
	\section{Introduction}
	It is commonplace in scientific research for investigators to rely on observational data to address questions of interest. While randomized experiments are the gold standard for drawing causal inferences about the effect of a treatment (also known as exposure, intervention, regime, or policy), in many cases, randomized experiments are difficult or infeasible to implement for logistical, financial or ethical reasons. For example, it would be unethical to force people to smoke to study the causal effect of smoking on health outcomes. Instead, researchers must utilize observational data and make careful corrections to address various biases. Undeniably, it is considerably more difficult to draw correct causal conclusions from observational data than from a randomized experiment. The main reason is due to confounding induced by non-randomization of treatments. Usually, researchers make unverifiable assumptions to draw causal conclusions of treatment effects, such as unconfoundedness of the treatment-outcome relationship, after adjusting for a set of confounders. Current causal inference methods, including propensity score methods \citep{rosenbaum1983central}, outcome regression methods, and doubly robust methods \citep{robins1995analysis,lunceford2004stratification,bang2005doubly,rotnitzky2015double}, have been developed to remove confounding bias, mainly in the settings where confounders are fully observed. However, observational data is also highly prone to missingness. Thus it is important, and at many times critical, to handle missing data properly to avoid introducing additional bias to the data analysis. 
	
	\subsection{Missing Data}
	Despite the best intentions of researchers, missing data is near impossible to avoid in real-world settings. Fortunately, as prevalent as missingness is, so too are methods with which to address missingness; however, the type of missingness matters when selecting a method.
	
	Missing data occurs by way of one of three mechanisms in observational data. The first and simplest type is Missingness Completely at Random (MCAR) \citep{rubin1976inference}. In this setting, whether or not an observation is missing is independent of both the observed and missing data. The second is Missingness at Random (MAR). Here, whether or not an observation is missing depends only on the observable data but not the missing data. Lastly, missing data can be Missing Not at Random (MNAR). In this setting, even after conditioning on all observed data, missingness will still depend on the missing portion of the data. 
	
	Confirming which missingness type is present for a given data set can not be validated from only the observed data. \cite{little1988MCAR_Test} proposed a chi-square based test capable of checking if the MCAR assumption was violated, and more recently Mohan and Pearl have been able to use m-graphs to refute instances where the MAR and MNAR assumptions were proposed but shown not to hold \citep{mohan2014testability,mohan2018graphical}. Despite not being able to validate if a particular missingness model is true, it is still common for arguments towards one of these assumptions to be made, relying on the knowledge of subject matter experts for the data at hand. If the observed data is sufficient to explain the missingness, the MAR assumption is plausible, and it is this assumption we carry forward for the majority of our discussion. In Section 6 we will discuss the extension to MNAR.
	
	\subsection{Approaches to Address Missing Data}
	Many methods have been proposed to address missingness when assuming a MAR pattern. Two of the most common approaches in practice are complete case estimation (CC) and multiple imputation (MI). In particular, MI was favored by the National Research Council in 2010 as one of its preferred means of addressing missing data in clinical trials \citep{NRC2010missingness}. Under CC, all records with missing data are excluded, and treatment effects are estimated only on fully observed cases. CC can be biased under MAR; more importantly, if multiple variables have missing values, there will only be a small portion of complete cases in the data. By throwing out a large portion of the data, the effective sample size shrinks, inflating variances. Therefore, CC suffers from loss of efficiency by utilizing less of the observed data in the final analysis \citep{WhiteCarlin2010}.
	
	MI, on the other hand, is traditionally recommended for MAR in part due to its improved efficiency over CC and due to it being applicable to a wider range of missingness mechanisms. With MI, the full joint distribution of the data is estimated (either empirically or modeled based on distributional assumptions), and from this, a series of $M$ new imputed data sets are drawn. In each imputed data set,  MI fills in each missing value with an imputed value by sampling from the posterior predictive distribution of the missing value given the observed values. Then, full sample analyses can be applied to each imputed data sets, and these multiple results are summarized by an easy-to-implement combining rule for inference \citep{rubin1987multiple}. 
	
	MI has produced valid frequentist inference in a wide range of applications \citep{Clogg1991mi_application}. At the same time, Rubin's variance estimator for MI has been shown to not always be consistent \citep{fay1992inferences, kott1995paradox, fay1996mi, bindersun1996mi, WangRobins1998mi, robinswang2000imputation, nielsen2003propermi, kim_et_al2006mi_bias, yang_kim2016mi}. For MI inference to be consistent, imputation must be proper (see \citet{rubin1987multiple} for a precise definition). Practically speaking, for any proper imputations, under sufficiently regular models, Rubin's combining rule provides a consistent estimator of the parameter of interest and a weekly unbiased estimator of its asymptotic variance.
    When the imputation model is correctly specified and the MAR assumption holds, \cite{meng1994multiple} showed that a sufficient condition for imputation to be proper is that the imputation model and the analysis model are \textit{congenial}. For example, the imputation model is correctly specified and the analysis is efficient under the same model. Congeniality is sometimes more elusive than it would appear. Even when the imputation model is correctly specified, \citet{yang_kim2016mi} showed that MI is not necessarily congenial for method of moments estimation. Therefore, some common statistical procedures may be incompatible with MI. From a causal inference perspective, this poses a problem as the validity of Rubin's variance estimator has not been fully explored for many full sample estimation methods used widely in causal inference. Certain otherwise unbiased and consistent full sample causal inference methods (outcome regression, weighting, matching, etc.) may lose these properties when applied in conjunction with MI and MI-produced data sets. Many of the most common estimates for average treatment effects are based on method of moments estimators and are thusly susceptible to inaccurate variance estimates when using Rubin's variance estimator for MI. For researchers desiring to make causal claims when utilizing MI, it is imperative for the variance properties of their estimators to, therefore, be either validated or an alternative method must be proposed. 
	
	\subsection{Fractional Imputation as an Alternative}
	As an alternative to CC and MI, there are likelihood-based methods that can be applied. When using these methods, the key insight is that under fully observed confounders, the full sample estimators are obtained by solving estimating equations. In the presence of partially observed confounders, the corresponding estimators can be obtained by solving conditional estimating equations which integrate out the missing confounders given the observed data. There are two difficulties in this approach. First, it requires consistent estimators in the conditional distribution of the missing confounders given the observed data, such as MLE. In the presence of missing values, an EM algorithm is typically used. Second, numerical integration is needed. Integration approximated by imputation was considered by many authors, such as Monte Carlo EM method s \citep{wei1990monte}. For Monte Carlo EM algorithm s, in each E-step, the imputed values are regenerated, and thus the computation can be quite heavy. Also, the convergence of Monte Carlo sequence of the estimators is not guaranteed for fixed Monte Carlo sample size \citep{booth1999maximizing}. 
	
	In practice, EM algorithms may not be feasible when the conditional expectation in the E-step is not available in a closed form. Instead, fractional imputation (FI) has been proposed to serve as a computational tool for implementing the expectation step (E-step) in the EM algorithm \citep{wei1990monte,kim2011parametric,yangkim2016review}, which simplifies computation by drawing on importance sampling to obtain the fractional weights and reducing the iterative computation burden over other simulation methods such as Markov Chain Monte Carlo. See \citet{Yang2013imputation,Yang2013parametric,Kim2014Fractionalhotdeck,Yang2015likelihood-based} for applications of FI outside of the causal inference context. The main idea in FI is to produce a complete data set by imputation and each imputed value is associated with a fractional weight, by which the observed likelihood can be approximated by the weighted average of the imputed data likelihood. The resulting estimator approximates the maximum likelihood estimator.

\begin{figure}[h]
	    \centering
	    \includegraphics[width=.8\linewidth]{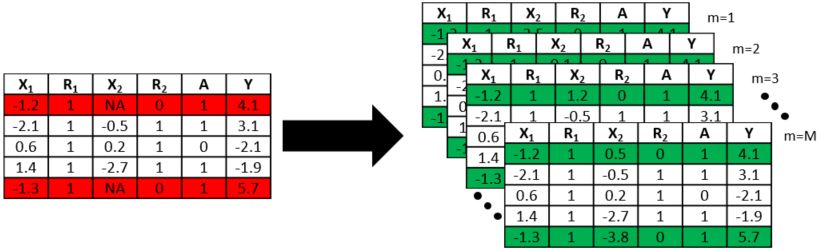}
	    \caption{Imputed Data via MI}
	    \label{fig:MI_picture}
	\end{figure}

	For illustrative purposes, we can compare the format of the imputed data via FI to the more commonly encountered imputed data via MI. Suppose we have a data set where some of the records are subject to missingness. MI will attempt to address this missingness by creating $M$ data sets where each imputed value is imputed exactly once in each of the $M$ data sets as depicted in Figure \ref{fig:MI_picture}. In this image $X_1$ is a fully observed covariate, and $A$ and $Y$ are fully observed treatment indicator and response variables respectively. $X_2$ is a covariate subject to missingness. $R_1$ and $R_2$ are missingness indicators for $X_1$ and $X_2$ where $R_{ij}=0$ indicates $X_j$ is missing for record $i$. Records with missingness are indicated in red on the left. The completed data sets (with $X_2$ imputed) are on the right with the records now including imputed values for $X_2$ highlighted in green.
	
\begin{figure}[h]
	    \centering
	    \includegraphics[width=.8\linewidth]{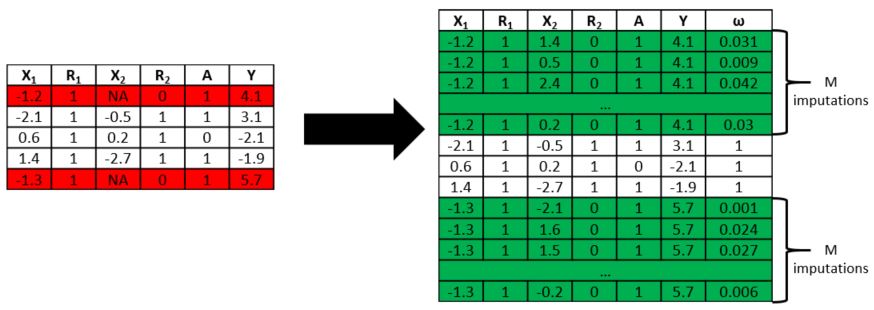}
	    \caption{Imputed Data via FI}
	    \label{fig:FI_picture}
	\end{figure}

	FI instead seeks to generate a single imputed data set, but one where each imputed record also includes a fractional weight. That fractional weight is indicative of how likely the imputed data is to occur under the distribution of the completed data set. Figure \ref{fig:FI_picture} shares all the same features as Figure \ref{fig:MI_picture} except now includes an additional column in the imputed data for the fractional weight ($\omega$) that will be incorporated into analyses.
	
	At its most basic level, the difference in imputation approach is analogous to a debate between a wide vs a tall data set. It could even be argued that when recombining data sets imputed via MI an implicit "weight" of $1/M$ is assigned to every record. This implicit $1/M$ weight acts similarly to the fractional weights $\omega$ generated via FI. The important distinction between MI and FI lies in how the final "weights" assigned to each imputed record are produced. The weight generation process for FI will be discussed in more depth in Section \ref{sec:FI_Details}. For the time being, we will assume these weights are generated appropriately and can be incorporated easily into further analyses.
	
	FI is sufficiently versatile that like MI it can be deployed along-side both continuous and categorical variables. Under the same common regularity conditions as seen often with MI, we show that the FI estimators are consistent and asymptotically normal. The remainder of this paper focuses on expanding FI into the causal literature by developing an FI-based method for estimating causal treatment effects. Once developed we will validate the method relative to existing causal inference methods. Specifically, we investigate the comparative performance of FI vs MI and CC for estimating causal treatment effects when confounders are subject to missingness.
	
	In summary, the proposed FI framework achieves desirable properties for causal inference:
	\begin{enumerate}
	    \item the same fractionally imputed data set allows for applying general full-sample estimators (which solve certain estimating equations) of the average treatment effect, including regression estimators, inverse probability of weighting estimators, and augmented weighting estimators;
	    \item the FI estimators are asymptotically linear and therefore allow resampling methods for variance estimation and inference, which is simple to implement in practice;
	    \item the unified FI inference has theoretical guarantees and offers a solution to the uncongeniality issue of MI;
	    \item and lastly, FI is not only statistically efficient but also computationally efficient compared to MI, as demonstrated via simulation and real-world application.
	\end{enumerate}
	
	The rest of the article is organized as follows. We begin in Section 2 with a description of notation and assumptions. In Section 3 we fully outline the FI process as well as derive the resulting variance estimator for the treatment effect estimator.  We implement a simulation study in Section 4 comparing the accuracy and precision of treatment effect estimators when the missingness is addressed by CC, MI, and FI. Section 5 provides a demonstration of FI's utilization with a real-world health data set. Finally, in Section 6, we end with a discussion of the results and of the implication they have on current and future causal work.
	
	\section{Setup and Notation}
	\subsection{Treatment Effect Estimation Notation}
	
	Following \citet{splawa1990application} and \citet{rubin1974estimating} we use the potential outcomes framework. Treatment is denoted by $A_{i}\in\{0,1\}$, where $0$ and $1$ are labels for control and treatment respectively. For each subject $i$, define a pair of potential outcomes $\{Y_{i}(1),Y_{i}(0)\}$ which represent the outcomes if the subject was treated, $Y_{i}(1)$, and if he or she was not, $Y_{i}(0)$. Implicit in this notation, we make the stable unit treatment value assumption \citep{rubin1978bayesian}. The observed outcome for subject $i$ is then  $Y_{i}=Y_{i}(A_{i})$. Let $X_{i}$ be the vector of confounders for subject $i$. We assume that $\{X_{i},A_{i},Y_{i}(1),Y_{i}(0)\}_{i=1}^{n}$ are independent draws from the distribution $\{X,A,Y(1),Y(0)\}$, and therefore, $\{(X_{i},A_{i},Y_{i})\}_{i=1}^{n}$ are independent and identically distributed. The conditional treatment effect is $\tau(X)=E\{Y(1)-Y(0)\mid X\}$, and the average treatment effect is $\tau_{0}=E\{\tau(X)\}$. The average treatment effect cannot be estimated without further assumptions, because for each subject only one potential outcome is observed. The common assumptions for identifying the average treatment effect \citep{rosenbaum1983central} are as follows: \vspace{3mm}
	
	\begin{flushleft}
	    \begin{assumption}[Ignorability]$Y(a)\indep A\mid X$ for $a=0,1$, where $\indep$ means "is conditionally independent of".
	    \end{assumption}
		\begin{assumption} [Sufficient overlap] With probability 1, $0<c_{1}\leq e(X)\leq c_{2}<1$, where $e(X)=\mathrm{pr}(A=1\mid X)$ is the propensity score. 
		\end{assumption}
	\end{flushleft}
	
	\vspace{3mm}
	
	Under Assumption 1, adjusting for covariates $X$ creates a randomization-like scenario and removes confounding biases brought on by treatment selection (i.e., for each level of $X$, the treatment assignment is as good as randomization). However, in practice, there are often many variables in $X$, some of which are continuous; therefore, directly conditioning on each level of $X$ is difficult. Alternatively, the propensity score has been proposed as a one-dimensional summary of $X$ \citep{rosenbaum1983central}. The central role of the propensity score lies in the fact that Assumption 1 implies $Y(a)\indep A\mid e(X)$ for $a=0,1$. Therefore, adjusting for the propensity score alone can remove confounding biases.  
	
	\subsection{Treatment Effect Estimation Under Fully Observed Data}
	Reliance on the propensity score comes about naturally when we decompose the joint density of $X$, $A$, and $Y$ into three particular components. Specifically
	\begin{align}
	f(X,A,Y)&=f(X)f(A|X)f(Y|X,A) \nonumber\\ 
	&=f(X)e(X)f(Y|X,A) \label{eq:simple_decomp}
	\end{align}
	Based on this decomposition, a number of propensity score based estimators have been proposed for estimating the treatment effect including propensity score matching, subclassification, or weighting. See \citet{imbens2015causal} for a textbook discussion. If we limit the class of propensity estimators to only parametric estimates (of the form $e(X\mid\theta)$ where $\theta$ is the vector of parameters used to estimate the propensity score), the two most common estimates of $\tau$ from this class are that of Inverse Propensity Weighting (IPW) and Augmented IPW (AIPW). Both are illustrated here as examples. In both examples, let $e(X\mid\hat{\theta})$ be the estimated propensity score where $\theta$ has been estimated by some consistent estimator $\hat{\theta}$. In practice, $e(X\mid\hat{\theta})$ is typically a logistic regression model.
	
	\begin{example}[IPW estimator]
		The IPW estimator of $\tau_{0}$ is 
		\begin{equation}
		\hat{\tau}_{IPW}=n^{-1}\sum_{i=1}^n\left\{\frac{A_iY_i}{e(X_i\mid\hat{\theta})}-\frac{(1-A_i)Y_i}{1-e(X_i\mid\hat{\theta})}\right\}.
		\end{equation}
	\end{example} 
	\begin{example} [AIPW estimator]
		Let $\mu(X,a\mid\hat{\beta})$ be an unbiased estimator of $E(Y\mid X,A=a;\beta)$, then the AIPW estimator of $\tau_{0}$ is 
		\begin{equation}
		\begin{aligned}
		\hat{\tau}_{AIPW}=n^{-1}\sum_{i=1}^n\left(\left[\frac{A_iY_i}{e(X_i\mid\hat{\theta})}+\left\{1-\frac{A_i}{e(X_i\mid\hat{\theta})}\right\}\mu(X_i,1\mid\hat{\beta})\right]-\right.\\ \left.\left[\frac{(1-A_i)Y_i}{1-e(X_i\mid\hat{\theta})}+\left\{1-\frac{1-A_i}{1-e(X_i\mid\hat{\theta})}\right\}\mu(X_i,0\mid\hat{\beta})\right]\right).
		\end{aligned}
		\end{equation}
	\end{example}
	
	We adopt the estimating equations convention where we let $U(\tau;X,A,Y\mid\eta)$ be the estimating function for $\tau_{0}$ under a given set of nuisance parameters $\eta$. An unbiased estimate for $\tau_{0}$ can then be derived as the solution to $P_{n}U(\tau;X,A,Y\mid\hat{\eta})=0$ where $\hat{\eta}$ is a consistent estimator of $\eta$ and $P_{n}$ is the empirical measure; namely $P_{n}f(X)=n^{-1}\sum_{i=1}^{n}f(X_i)$. As examples, estimating functions for IPW and AIPW are shown below.
	
	\begin{example}[IPW estimating function]
		The estimating function of $\hat{\tau}_{IPW}$ is
		$$ U_{IPW}(\tau;X,A,Y\mid\hat{\eta})=\frac{AY}{e(X\mid\hat{\theta})}-\frac{(1-A)Y}{1-e(X\mid\hat{\theta})}-\tau.$$
		Here $\hat{\eta}=(\hat{\theta})$, the parameter estimates used to calculate the propensity scores.
	\end{example}
	
	\begin{example}[AIPW estimating function]
		The estimating function for $\hat{\tau}_{AIPW}$ is
		\begin{multline*}
		U_{\mathrm{AIPW}}(\tau;X,A,Y\mid\hat{\eta})=\left[\frac{AY}{e(X\mid\hat{\theta})}+\left\{ 1-\frac{A}{e(X\mid\hat{\theta})}\right\} \mu(X,1\mid\hat{\beta})\right]\\	-\left[\frac{(1-A)Y}{1-e(X\mid\hat{\theta})}+\left\{ 1-\frac{1-A}{1-e(X\mid\hat{\theta})}\right\} \mu(X,0\mid\hat{\beta})\right]-\tau.
		\end{multline*}
		Here $\hat{\eta}=(\hat{\beta},\hat{\theta})$, the parameter estimates used to calculate $\mu(X,a\mid\hat{\beta})$ and the propensity scores respectively.
	\end{example}
	
	\begin{remark}[Doubly Robust Estimation]\label{Rmk:DR_FullData} The IPW estimate in Example 3 does not need to model the outcome $Y$, but it does require a correct model for $e(X\mid\theta)$. On the other hand, the AIPW estimate for $\tau_{0}$ obtained from the mean estimating function in Example 4, incorporates a double robust (DR) feature for estimation. That is, ${U}_{\mathrm{AIPW}}$ is an unbiased estimating function for $\tau_0$ if either $e(X\mid\theta)$ or $\mu(a,X\mid\beta)$ is correctly specified. 
	\end{remark} 
	
	\subsection{Missing Data Notation}
	Whereas all of the above discussion holds under fully observed responses and confounders, in this article we consider the case where $X$ contains missing values. To that end, let $R$ be a collection of indicator variables $R=(R_{1},...,R_{p})$ corresponding to $(X_1,...,X_p)$ where $R_{ij}=1$ indicates that $X_j$ is observed for subject $i$ and $R_{ij}=0$ indicates $X_j$ is not observed for subject $i$. 
	Let $\mathcal{R}$ denote the collection of all possible missingness patterns.
	Let $X_{\mathrm{obs,i}}$, the observed part of covariates for individual $i$, consist of $X_{ij}$ with $R_{ij}=1$. Similarly let $X_{\mathrm{mis,i}}$, the missing part of covariates for individual $i$, consist of $X_{ij}$ with $R_{ij}=0$. Note that the dimension of $ X_{\mathrm{obs,i}}$ and $X_{\mathrm{mis,i}}$ will vary across individuals. For notational simplicity, we suppress the subscript $i$ for subject.
	
	If we return to equation (\ref{eq:simple_decomp}) and incorporate the parameterizations laid out in examples 1 and 2, the decomposition of the joint distribution can be naturally extended to account for missingness. To do so, we rewrite the joint distribution as:
	
	\begin{align}
	f(X,A,Y,R;\alpha,\beta,\theta,\rho) &=f(X)e(X|\theta)f(Y|X,A;\beta)f(R|X,A,Y;\rho)\nonumber\\
	&=f(X_{\mathrm{obs}},X_{\mathrm{mis}})e(X|\theta)f(Y|X,A;\beta)f(R|X,A,Y;\rho)\nonumber\\	
	&=f(X_{\mathrm{obs}})f(X_{\mathrm{mis}}|X_{\mathrm{obs}};\alpha)e(X|\theta)f(Y|X,A;\beta)f(R|X,A,Y;\rho) \label{eq:full_decomp}
	\end{align}
	where $\alpha$ is the collection of parameters used to estimate the missing portion of $X$ given the observed portion of $X$, and $\rho$ is the collection of parameters used in describing the missingness mechanism.
	
	Finally, in this article we are only interested in the case where $X_{\mathrm{mis}}$ follows a MAR pattern, which leads us to our final assumption which completes the basis for Theorem \ref{Thm:IPTW} (the proof for which is made available in the appendix).
	\begin{assumption}[Missing at random] $\mathcal{R}\indep X_{\mathrm{mis}}\mid X_{\mathrm{obs}},A,Y$
	\end{assumption}
	
	Here we adopt the notion of Rubin's MAR in the sense that MAR may hold for different variables depending on the missingness pattern. The scientific justification of this assumption may be difficult; however, theoretically it is the weakest condition under which the missingness process can be ignored \citep{rubin1976inference}. Alternatively, one can consider the variable-based taxonomy of MAR \citep{mohan2018graphical}, where $X_{\mathrm{mis}}$ represents variables that are subject to missingness and $X_{\mathrm{obs}}$ represents variables that are fully observed. Our method development is similar under this notion of MAR but using the new definition of $X_{\mathrm{mis}}$ and $X_{\mathrm{obs}}$ in equation (\ref{eq:full_decomp}).
	
	\begin{theorem}\label{Thm:IPTW}Under Assumptions 1\textendash 3, $\tau_{0}$ is identifiable from the observed data. 
	\end{theorem} 
	
	A convenient consequence of adopting an MAR framework is that because of Assumption 3, any expectations taken with respect to $X_\mathrm{mis}$ will result in $f(R|X,A,$
	$Y;\rho)$ falling out of the decomposition; therefore for the remainder of this article, unless otherwise noted, we will be suppressing inclusion of an explicit missingness mechanism term and instead using the decomposition: 
	
	\begin{equation}
	f(X,A,Y;\alpha,\beta,\theta) =f(X_{\mathrm{obs}})f(X_{\mathrm{mis}}|X_{\mathrm{obs}};\alpha)e(X|\theta)f(Y|X,A;\beta) \label{eq:MAR_decomp}
	\end{equation}
	
	\subsection{Treatment Effect Estimation Under MAR}
	In the presence of missingness, we can let $\overline{U}(\tau;X_{\mathrm{obs}},A,Y\mid\hat{\eta})=E\{U(\tau;X,A,Y\mid\hat{\eta})\mid X_{\mathrm{obs},}A,Y\}$. We term $\overline{U}(\tau;X_{\mathrm{obs}},A,Y\mid\hat{\eta})$ the "mean estimating function" of $\tau_{0}$ given the observed data. Note that in defining $\overline{U}(\tau;X_{\mathrm{obs}},A,Y\mid\hat{\eta})$, $\hat{\eta}$ must now also expand to include any new parameters $\hat{\alpha}$ utilized in estimating $\mathrm{E}(X_{\mathrm{mis}}\mid X_{\mathrm{obs}},A,Y;\alpha)$. From Theorem \ref{Thm:IPTW}, under Assumptions 1\textendash 3, $\overline{U}(\tau;X_{\mathrm{obs}},A,Y\mid\hat{\eta})$ is an unbiased estimating function of $\tau_{0}$. Therefore, a consistent estimator of $\tau_{0}$ can be obtained by solving $P_{n}\overline{U}(\tau;X_{\mathrm{obs}},A,Y\mid\hat{\eta})=0$.
	
	\begin{remark}[Doubly Robust Estimation Under MAR]\label{Rmk:estimation2} As with IPW estimation under fully observed data, IPW estimation under MAR does not need to model the outcome $Y$, but it does now require correct models for both $e(X\mid\theta)$ and $f(X_{\mathrm{mis}}\mid X_{\mathrm{obs}},A,Y;\alpha)$. On the other hand, the AIPW estimate for $\tau_{0}$ will still be unbiased provided $f(X_{\mathrm{mis}}\mid X_{\mathrm{obs}},A,Y;\alpha)$ is correctly specified and either $e(X\mid\theta)$ or $\mu(a,X\mid\beta)$ is correctly specified. The DR feature of AIPW estimation for treatment effects has been shown extensively in full data situations and more recently in the case where exposures are MAR \citep{Zhang2013triplerobust}.
	\end{remark} 
	
	\section{Fractional Imputation } \label{sec:FI_Details}
	Estimation of treatment effects under fully observed data is straightforward; unfortunately, fully observed data is rarely encountered in practice. Imputation methods are often used to facilitate estimation in the presence of missing values by completing the partially observed portions of the data and coaxing a "full" data set out of a partial one. It is important to note, though, imputation methods only provide a means of addressing the missingness and complete the data set without heed given to effect estimation. Therefore, to examine the consequences of choosing a specific imputation method, we propose a two-stage procedure. In the first stage, referred to as the design stage, we use an imputation technique to fill in the missing covariate values and estimate the propensity scores. In the second stage, referred to as the analysis stage, classical propensity score techniques are applied to estimate the causal parameters. See \citet{rubin2007design}, \citet{rubin2008objective}, and \citet{stuart2010matching} for the mention of decomposing causal inference into two different stages. 
	
	This framework will be used for both FI and MI methods. The former of these methods we discuss in more detail here. For a more detailed examination of MI see \citet{rubin1987multiple}.
	
	\subsection{Implementing Fractional Imputation }
	
	Earlier in this paper, Figure \ref{fig:FI_picture} depicted what a completed data set might look like after using FI to impute missing data during the design stage. Besides the shape of the data, the addition of an explicit weight to each record of the imputed data set stands out from more commonly encountered MI data sets. To generate fractional weights, FI deploys a three-step process. First, missing values are imputed from some proposed distribution, then fractional weights are updated, then model parameters for the full joint distribution are re-estimated (now under updated weights). The re-weighting and model updating steps cycle until convergence.
	
	In the first step, sometimes referred to as the imputation step or the I-step \citep{kim2013statistical,yangkim2016review}, every missing value is imputed $M$ times by means of some proposal function $h(X_{mis}\mid X_{obs})$.	This generates $M$ new values $X_{mis,i}^{*(j)}$ for each partially observed record. The choice of $h()$ is arbitrary and left to the imputer, but a convenient choice is $f(X_{mis}\mid X_{obs}; \alpha)$ to align with the decomposition from equation \ref{eq:MAR_decomp}. This choice would necessitate being able to provide or estimate a value for $\alpha$ (for instance $\alpha=\hat{\alpha}_0$ where $\hat{\alpha}_0$ is the MLE for $\alpha$ calculated only using complete cases).
	
	At this first step, every record where a value was imputed will have a fractional weight of $\omega_{ij}^*=1/M$. Note, at the conclusion of every step in FI, fractional weights for each individual, observation, etc. $i$ are held to the condition $\sum_{j=1}^{M} \omega_{ij}^*=1$.
	
	In the second step, the weighting step or W-step, the fractional weights are updated proportional to the likelihood of the imputed value under the full joint distribution, divided by the likelihood of the imputed value under the proposal distribution $h()$ from the I-step. If choosing the decomposition from equation \ref{eq:MAR_decomp} that would appear as
	
	\begin{equation*}
	    \omega_{ij}^{*(t)}\propto\frac{f(X_{obs,i},X_{mis,i}^{*(j)},A_i,Y_i;\hat{\eta}^{(t)})}{h(X_{mis,i}^{*(j)})},
	\end{equation*}
	given the current parameter estimates for $\hat{\eta}^{(t)}=(\hat{\alpha}^{(t)},\hat{\beta}^{(t)},\hat{\theta}^{(t)})$. 
	
	In the third step, the maximization step or M-step, the parameter values used to estimate the full joint distribution are updated given the new values of $\omega_{ij}^*$ from $\hat{\eta}^{(t)}$ to $\hat{\eta}^{(t+1)}$. The W-step and M-step iterate, setting t=t+1 after each M-step, until the model parameters converge. The resulting data set with the final fractional weights included can then be passed on to the analysis stage as if it were a complete data set (similar to MI passing on $M$ data sets). An example generating a complete data set via FI is included in the appendix.
	
	\subsection{Characterizing Fractional Imputation for Estimating Treatment Effects}
	
	For illustration, consider the case where $X$ contains only two variables, $X=(X_{1},X_{2})$, where $X_{1}$ is fully observed and $X_{2}$ is subject to missingness. Let $R_2$ be the response indicator of $X_{2}$. From Examples 3 and 4, we can obtain an estimator of $\tau_{0}$ by solving the mean estimating equation
	\begin{multline}
	\sum_{i=1}^{n}\overline{U}(\tau;X_{\mathrm{obs},i},A_{i},Y_{i}\mid\hat{\eta})=\sum_{i=1}^{n}E\{U(\tau;X_{i},A_{i},Y_{i}\mid\hat{\eta})\mid X_{\mathrm{obs},i},A_{i},Y_{i}\}\\
	=\sum_{i=1}^{n}R_{2i}U(\tau;X_{i},A_{i},Y_{i}\mid\hat{\eta})+\sum_{i=1}^{n}(1-R_{2i})E\{U(\tau;X_{i},A_{i},Y_{i}\mid\hat{\eta})\mid X_{1i},A_{i},Y_{i}\}=0,\label{eq:conditionalEE}
	\end{multline}
	where $U(\tau;X_i,A_i,Y_i\mid\hat{\eta})$ denotes either $U_{IPW}(\tau;X_i,A_i,Y_i\mid\hat{\eta})$ or $U_{AIPW}(\tau;X_i,A_i,Y_i\mid\hat{\eta})$.
	
	In (\ref{eq:conditionalEE}), the conditional expectation $E\{U(\tau;X_{i},A_{i},Y_{i}\mid\hat{\eta})\mid X_{\mathrm{obs},i},A_{i},Y_{i}\}$ is often difficult to obtain. The basic idea of FI is to overcome this difficulty by creating a weighted set $\{(\omega_{ij}^*,X_i^{*(j)},A_i,Y_i):j=1,...,M\}$ such that $E\{U(\tau;X_{i},A_{i},Y_{i}\mid\hat{\eta})\mid X_{\mathrm{obs},i},A_{i},Y_{i}\}$ can be approximated by $\sum_{i=1}^n\sum_{j=1}^M\omega_{ij}^*U(\tau;X_{i}^{*(j)},A_{i},Y_{i}\mid\hat{\eta})$. 
	
	\begin{remark}
		Only records where $R_{2i}=0$ need imputed, so when utilizing FI, only these observations require a weight $\omega_{ij}^*$ to be calculated in the weighted set $\{(\omega_{ij}^*$,
		$X_i^{*(j)},A_i,Y_i):j=1,...,M\}$. However, implicit in this representation is the generation of weights $\omega_i=1$ for observations where $R_{2i}=1$. While notation for such implicit generation is suppressed in this article, if desired, the weighted set can be rewritten as $\{(\omega_{ij}^*,X_i^{*(j)},A_i,Y_i):i=1,...,n; j=1,...,m_i\}$ where
		\begin{center}
			\begin{minipage}{.3\linewidth}
				\begin{center}
					$m_i=
					\begin{cases}
					M & $if $R_{2i}=0 \\
					1 & $if $R_{2i}=1
					\end{cases},$
				\end{center}
			\end{minipage}%
			and
			\begin{minipage}{.45\linewidth}
				\begin{center}$X_i^{*(j)}=
					\begin{cases}
					(X_{obs,i},x_i^{*(j)}) & $if $R_{2i}=0 \\
					(X_{obs,i}) & $if $R_{2i}=1
					\end{cases}.$
				\end{center}
			\end{minipage}
		\end{center}

		\noindent Such notation may be beneficial if equational symmetry is desired, though the FI process and resulting estimates of $\tau_{0}$ are unaffected.
		\label{Rem:altWeightNotation}
	\end{remark}

	Toward that goal of approximating $E\{U(\tau;X_{i},A_{i},Y_{i}\mid\hat{\eta})\mid X_{\mathrm{obs},i},A_{i},Y_{i}\}$ as: \newline \noindent
	$\sum_{i=1}^n\sum_{j=1}^M\omega_{ij}^*U(\tau;X_{i}^{*(j)},A_{i},Y_{i}\mid\hat{\eta})$, notice that the last conditional expectation in (\ref{eq:conditionalEE}) can be written as
	\[
	E\{U(\tau;X_{i},A_{i},Y_{i}\mid\hat{\eta})\mid X_{1i},A_{i},Y_{i}\}=\frac{\int U(\tau;X_{1i},x_{2},A_{i},Y_{i}\mid\hat{\eta})f(X_{1i},x_{2},A_{i},Y_{i}\mid\hat{\eta})dx_{2}}{\int f(X_{1i},x_{2},A_{i},Y_{i}\mid\hat{\eta})dx_{2}},
	\]
	where $f(X_{1},X_{2},A,Y\mid\hat{\eta})$ is the joint distribution of $(X_{1},X_{2},A,Y)$ with nuisance parameters $\eta$ set to $\hat{\eta}$. Furthermore, the joint distribution can be decomposed similar as in (\ref{eq:MAR_decomp}) to be 
	\begin{multline}
	f(X_{1},X_{2},A,Y\mid\hat{\eta})=f(X_{1},X_{2}\mid\hat{\alpha})f(A\mid X_{1},X_{2};\hat{\theta})f(Y\mid X_{1},X_{2},A;\hat{\beta})\\=f(X_{1})f(X_{2}\mid X_{1};\hat{\alpha})e(X\mid\hat{\theta})f(Y\mid X_{1},X_{2},A;\hat{\beta}),
	\end{multline}
	where we assume $f(X_{2}\mid X_{1};\hat{\alpha})$ and $f(Y\mid X_{1},X_{2},A;\hat{\beta})$ be correctly specified as $f(X_{2}\mid X_{1};\alpha)$ and $f(Y\mid X_{1},X_{2},A;\beta)$, respectively. 
	
	Under complete response, the maximum likelihood estimator (MLE) of $\theta$ can be obtained as a solution to the score equations,
	\[
	\sum_{i=1}^{n}S(\theta;X_{i},A_{i})=0,
	\]
	where $S(\theta;X,A)$ is the score function of $\theta$ and can be written as  $S(\theta;X,A)=\partial\log f(A\mid X;\theta)/\partial\theta$ with $f(A\mid X;\theta)=e(X\mid\theta)^{A}\{1-e(X\mid\theta)\}^{1-A}$  \citep{louis1982finding,pfeffermann1998weighting}, which under missingness is rewritten
	\begin{equation}
	\sum_{i=1}^{n}R_{2i}S(\theta;X_{i},A_{i})+\sum_{i=1}^{n}(1-R_{2i})E\{S(\theta;X_{i},A_{i})\mid X_{1i},A_{i},Y_{i}\}=0.\label{eq:conditionalEE-theta}
	\end{equation}
	MLE estimates of $\hat{\alpha}$ and $\hat{\beta}$ can be obtained similarly to (\ref{eq:conditionalEE-theta}) for their respective mean score equations. 
	
	For $\hat{\alpha}$ the mean score equation is
	\begin{equation}
	\sum_{i=1}^{n}S(\alpha;X_{i})=\sum_{i=1}^{n}S(\alpha;X_{1i},X_{2i})=0, \label{eq:conditionalEE-alpha}
	\end{equation}
	where $S(\alpha;X)$ is the score equation for $\alpha$ written as $S(\alpha;X)=S(\alpha;X_1,X_2)=\partial\log f(X_2\mid X_1;\alpha)/\partial\alpha$. Under MAR,  $\hat{\alpha}$ can be obtained using only complete cases. 
	
	For $\hat{\beta}$ the mean score equation is
	\[
	\sum_{i=1}^{n}S(\beta;X_{i},A_{i},Y_{i})=0,
	\]
	where $S(\beta;X,A,Y)$ is the score function of $\beta$ and can be written as $S(\beta;X,A,Y)=\partial\log f(Y\mid X,A;\beta)/\partial\beta$, which under missingness is rewritten
	\begin{equation}
	\sum_{i=1}^{n}R_{2i}S(\beta;X_{i},A_{i},Y_{i})+\sum_{i=1}^{n}(1-R_{2i})E\{S(\beta;X_{i},A_{i},Y_{i})\mid X_{1i},A_{i},Y_{i}\}=0.\label{eq:conditionalEE-beta}
	\end{equation}
	
	To obtain the solution to (\ref{eq:conditionalEE}), (\ref{eq:conditionalEE-theta}), (\ref{eq:conditionalEE-alpha})and (\ref{eq:conditionalEE-beta}), the EM algorithm can be applied. To do so using FI, the following process can be implemented: 
	\begin{description}
		\item [{Step 0.}] Let the initial values for parameters be set to $\alpha^{(0)}$, $\beta^{(0)}$, and $\theta^{(0)}$ which are the MLE of $\alpha$, $\beta$, and $\theta$ using only complete cases. For each unit $i$ with $R_{2i}=0$, generate $M$ imputed values of $X_{2i}$, denoted by $x_{2i}^{*(j)}$ $(j=1,\ldots,M)$, from a proposal distribution $h(x_{2})$, e.g. $f(x_{2}\mid X_{1};\alpha^{(0)})$.
		\item [{Step 1.}] At the $t^{th}$ EM iteration, compute the fractional weight
		\[
		\omega_{ij}^{*(t)}\propto\frac{f(x_{2i}^{*(j)}\mid X_{1i};\alpha^{(t)})e\left\{(X_{1i},x_{2i}^{*(j)})\mid\theta^{(t)}\right\}f(Y_{i}\mid A_{i},X_{1i},x_{2i}^{*(j)};\beta^{(t)})}{h(x_{2i}^{*(j)})}
		\]
		subject to $\sum_{j=1}^{M}\omega_{ij}^{*(t)}=1$. 
		\item [{Step 2.}] Use $\omega_{ij}^{*(t)}$ and $(X_{1i},x_{2i}^{*(j)},A_{i},Y_{i})$ to update the parameters from $(\alpha^{(t)},\beta^{(t)},\theta^{(t)})$ to $(\alpha^{(t+1)},\beta^{(t+1)},\theta^{(t+1)})$ by solving the respective imputed score equations.
		To update $\alpha^{(t)}$ to $\alpha^{(t+1)}$, solve the imputed score equation
		\[
		\sum_{i=1}^{n}R_{2i}S\left(\alpha;X_{1i},X_{2i}\right)+\sum_{i=1}^{n}(1-R_{2i})\sum_{j=1}^{M}\omega_{ij}^{*(t)}S\left(\alpha;X_{1i},x_{2i}^{*(j)}\right)=0.
		\]
		
		To update $\beta^{(t)}$ to $\beta^{(t+1)}$, solve the imputed
		score equation 
		\[
		\sum_{i=1}^{n}R_{2i}S(\beta;X_{1i},X_{2i},A_{i},Y_{i})+\sum_{i=1}^{n}(1-R_{2i})\sum_{j=1}^{M}\omega_{ij}^{*(t)}S(\beta;X_{1i},x_{2i}^{*(j)},A_{i},Y_{i})=0.
		\]
		To update $\theta^{(t)}$ to $\theta^{(t+1)}$, solve the imputed
		score equation 
		\[
		\sum_{i=1}^{n}R_{2i}S(\theta;X_{i},A_{i})+\sum_{i=1}^{n}(1-R_{2i})\sum_{j=1}^{M}\omega_{ij}^{*(t)}S(\theta;X_{1i},x_{2i}^{*(j)},A_{i})=0.
		\]
		\item [{Step 3.}] Set $t=t+1$ and go to Step 1. Continue until convergence. 
	\end{description}

	\begin{remark}
		Recall under MAR $\hat{\alpha}$ can be obtained under only complete cases. In such case, $\alpha^{(0)}=\hat{\alpha}$, and there is no need to update the parameter estimate $\alpha^{(t)}$ each iteration. Additionally, if $h(x_{2})=f(x_{2}\mid X_{1};\hat{\alpha})$, the calculation of the weight function simplifies to $\omega_{ij}^{*(t)}\propto e\left\{(X_{1i},x_{2i}^{*(j)})^T\theta^{(t)}\right\}f(Y_{i}\mid X_{1i},x_{2i}^{*(j)},A_{i};\beta^{(t)})$ and $\sum_{j=1}^{M}\omega_{ij}^{*(t)}=1$. The simplification is not necessary under MAR, but we mention it here for the event when additional computational resource efficiencies are desired for a particular application.
		\label{Rem:alphaMLEalternative}
	\end{remark}
	
	Let $\hat{\eta}=(\hat{\alpha},\hat{\beta},\hat{\theta})$ be the resulting estimates for the nuisance parameters. Note that at each EM iteration, imputed values of $X_2$ are not changed; only fractional weights are updated for each iteration. The weights $\omega_{ij}^{*}$, obtained at the end of the EM iteration, assigned to imputed values can be called fractional weights. The fractional weight represents a similarity measure between the imputed value and the missing value. 
	
	By incorporating these weights, the conditional estimating equation for $\tau_O$ can be approximated
	by 
	\[
	\sum_{i=1}^{n}R_{2i}U(\tau;X_{i},A_{i},Y_{i}|\hat{\eta})+\sum_{i=1}^{n}(1-R_{2i})\sum_{j=1}^{M}\omega_{ij}^{*}U(\tau;X_{1i},x_{2i}^{*(j)},A_{i},Y_{i}|\hat{\eta})=0,
	\]
	and $\hat{\tau}$ can be obtained by solving this imputed estimating
	equation for $\tau$. Here, $U(\tau;X,A,Y|\hat{\eta})$ can be either the IPW or AIPW estimating function.

	\subsection{Asymptotic Results}
	Because $\hat{\tau}$ is obtained through the method of estimating equations, we establish the asymptotic properties of $\hat{\tau}$, in a manner similar to  \citet{robinswang2000imputation}.
	\begin{theorem}
		Let $\eta=(\alpha,\beta,\theta)$ be a vector of the nuisance parameters, and let $\hat{\eta}=(\hat{\alpha},\hat{\beta},\hat{\theta})$ be the vector of corresponding MLE estimators converging in probability to $\eta_0=(\alpha_0,\beta_0,\theta_0)$, the true values of the nuisance parameters. Under certain regularity conditions, the solutions to (\ref{eq:conditionalEE}), $\hat{\tau}$, is consistent for $\tau_{0}$ and satisfies
		\begin{center}
			$\sqrt{n}(\hat{\tau}-\tau_{0})\rightarrow\mathcal{N}(0,V)$,
		\end{center}
		as $n\rightarrow \infty$, where 
		\begin{align*}
		\mathrm{V} & =  \lambda^{-1}\varOmega \lambda^{-1T},\\
		\lambda & =  \mathrm{E}\left\{\frac{\partial}{\partial\tau}U(\tau_0;X,A,Y)\right\},\\
		\varOmega & =  \mathrm{Var}\{\overline{U}(\tau_0;X_{\mathrm{obs}},A,Y,\eta_0)+\kappa S_{\mathrm{obs}}(\eta_0;X_{\mathrm{obs}},A,Y)\},\\
		\overline{U}(\tau;X_{\mathrm{obs}},A,Y,\eta) & =  \mathrm{E}\left\{U(\tau;X,A,Y)\mid X_\mathrm{obs},A,Y,\eta \right\},\\
		\kappa & =  \mathrm{E}\{U(\tau_0;X,A,Y)S_{\mathrm{mis}}^T(\eta_0;X_{\mathrm{obs}},A,Y)\}\mathcal{I}_{\mathrm{obs}}^{-1},\\
		S_{\mathrm{mis}}(\eta;X_{\mathrm{obs}},A,Y) & = \mathrm{E}\left\{\frac{\partial}{\partial\eta}\mathrm{log}f(X_\mathrm{mis}\mid X_\mathrm{obs},A,Y;\eta)\mid X_{\mathrm{obs}},A,Y;\eta\right\},\\
		S_{\mathrm{obs}}(\eta;X_{\mathrm{obs}},A,Y) & =  \mathrm{E}\left\{S\left(\eta;X,A,Y\right)\mid X_{\mathrm{obs}},A,Y;\eta\right\},\\
		S(\eta;X,A,Y) & =  \frac{\partial}{\partial\eta}\mathrm{log}f(X,A,Y;\eta),
		\end{align*}
		and 
		\begin{align*}
		\mathcal{I}_{\mathrm{obs}}&=-\mathrm{E}\left\{\frac{\partial}{\partial\eta^T}S_{\mathrm{obs}}(\eta_0;X_{\mathrm{obs}},A,Y)\right\}	\\
		&=-\mathrm{E}\left\{\frac{\partial}{\partial\eta^T}S(\eta_0;X,A,Y)\right\}+\mathrm{E}\left\{S_\mathrm{mis}(\eta_0;X_\mathrm{obs},A,Y)^{\otimes^2}\right\}
		\end{align*}
		where $B^{\otimes^2}\equiv BB^T$ for some matrix $B$.
		
		\label{Thm:ExactVarEst}
	\end{theorem}

	\begin{proof*}
		Let $\overline{U}(\tau\mid\eta)\equiv n^{-1}\sum_{i=1}^{n}\mathrm{E}\left\{U(\tau;X_i,A_i,Y_i)\mid X_{\mathrm{obs},i},A_i,Y_i,\eta\right\}$. Note that $\hat{\tau}$ and $\hat{\eta}$ satisfy $\overline{U}(\hat{\tau}\mid\hat{\eta})=0$ as $M\rightarrow \infty$. By use of Taylor expansions we can study the asymptotic properties of $\hat{\tau}$.
		
		First, by a Taylor expansion of $\overline{U}(\tau\mid\hat{\eta})$ about $\hat{\eta}=\eta_{0}$ we obtain
		\begin{equation}
		\overline{U}(\tau\mid\hat{\eta})=\overline{U}(\tau|\eta_{0})+\mathrm{E}\left\{\frac{\partial}{\partial\eta^T}\overline{U}(\tau\mid\eta_{0})\right\}(\hat{\eta}-\eta_0)+o_p\left(n^{-1}\right). \label{pf:TaylorEta1}
		\end{equation}
		Because $\overline{U}(\tau|\eta)=n^{-1}\int U(\tau;X_i,A_i,Y_i)f(X_{\mathrm{mis,i}}|X_{\mathrm{obs,i}},A_i,Y_i;\eta)\mathrm{d}X_{\mathrm{mis}}$
		we obtain 
		\begin{align}
		\mathrm{E}\left\{\frac{\partial}{\partial\eta^T}\overline{U}(\tau|\eta)\right\}&= \mathrm{E}\left\{n^{-1}\sum_{i=1}^{n}\int U(\tau;X_i,A_i,Y_i)\frac{\partial f(X_{\mathrm{mis,i}}|X_{\mathrm{obs,i}},A_i,Y_i;\eta)}{\partial\eta}\mathrm{d}X_{\mathrm{mis}}\right\}		 \nonumber \\
		&= \mathrm{E}\Bigg\{n^{-1}\sum_{i=1}^{n}\int U(\tau;X_i,A_i,Y_i)\frac{\partial \mathrm{log}f(X_{\mathrm{mis,i}}|X_{\mathrm{obs,i}},A_i,Y_i;\eta)}{\partial\eta}
		\nonumber\\&\hspace{2.5cm} f(X_{\mathrm{mis,i}}|X_{\mathrm{obs,i}},A_i,Y_i;\eta)\mathrm{d}X_{\mathrm{mis}}\Bigg\}			 \nonumber \\
		&= \mathrm{E}\left\{U\left(\tau;X,A,Y\right)S_{\mathrm{mis}}^T\left(\eta;X_\mathrm{obs},A,Y\right)\right\}. \label{pf:eq:PartialUExpectation}
		\end{align}
		To express $\hat{\eta}-\eta_{0}$ from (\ref{pf:TaylorEta1}) further, we note that the EM algorithm leads to the MLE of $\eta_{0}$ and therefore $\hat{\eta}$ satisfies
		\begin{equation}
		n^{-1}\sum_{i=1}^{n}\mathrm{E}\left\{S(\hat{\eta};X_i,A_i,Y_i)\mid X_\mathrm{obs,i},A_i,Y_i;\hat{\eta}\right\}=0 \label{pf:eq:EtaMLERational},
		\end{equation}
		which depends on $\hat{\eta}$ in two places, namely $S(\hat{\eta};X_i,A_i,Y_i)$ and the conditional expectation taken with respect to $f(X_\mathrm{mis}\mid X_\mathrm{obs},A,Y;\hat{\eta})$. Applying another Taylor expansion about $\hat{\eta}=\eta_{0}$ this time in (\ref{pf:eq:EtaMLERational}) leads to
		\begin{align}
		0=&n^{-1}\sum_{i=1}^{n}\mathrm{E}\left\{S(\eta_{0};X_i,A_i,Y_i)\mid X_\mathrm{obs,i},A_i,Y_i;\eta_{0}\right\}\nonumber\\
		&+n^{-1}\sum_{i=1}^{n}\left[\mathrm{E}\left\{\frac{\partial}{\partial\eta^T}S(\eta_{0};X_i,A_i,Y_i)\mid X_\mathrm{obs,i},A_i,Y_i;\eta_{0}\right\}\right.\nonumber\\
		&\hspace{10mm}\left.+\mathrm{E}\left\{S(\eta_{0};X_i,A_i,Y_i)S_\mathrm{mis}^T(\eta_{0},X_i,A_i,Y_i)\mid X_\mathrm{obs,i},A_i,Y_i;\eta_{0}\right\}\right]\nonumber\\
		&+o_p\left(n^{-1/2}\right).
		\end{align}
		Therefore we can express
		\begin{equation}
		\hat{\eta}-\eta_{0}\cong\mathcal{I}_{\mathrm{obs}}^{-1}n^{-1}\sum_{i=1}^{n}S_{\mathrm{obs}}(\eta_{0};X_\mathrm{obs,i},A_i,Y_i). \label{pf:eq:TaylorEtaDiff}
		\end{equation}
		Combining (\ref{pf:eq:PartialUExpectation}) and (\ref{pf:eq:TaylorEtaDiff}), ignoring the small order terms, (\ref{pf:TaylorEta1}) can be expressed in a linear form:
		\begin{align*}
		\overline{U_l}(\tau\mid\eta_{0})&=n^{-1}\sum_{i=1}^{n}\left[\overline{U}(\tau;X_\mathrm{obs,i},A_i,Y_i;\eta_{0})\right.\nonumber\\
		&\hspace{12mm}+\left.\mathrm{E}\left\{\mathrm{U}(\tau;X,A,Y)S_{\mathrm{mis}}^T\left(\eta;X_\mathrm{obs,i},A_i,Y_i\right)\right\}\mathcal{I}_{obs}^{-1}S_{obs}(\eta_0;X_\mathrm{obs,i},A_i,Y_i)\right]\\
		&=n^{-1}\sum_{i=1}^{n}\left\{\overline{U}(\tau\mid X_\mathrm{obs,i},A_i,Y_i;\eta_{0})+\kappa S_{obs}(\eta_0;X_\mathrm{obs,i},A_i,Y_i)\right\}.
		\end{align*} 
		with the $l$ being used to denote the linearization.
		
		Second, note that we now have $\overline{U}(\hat{\tau}\mid\hat{\eta})=\overline{U_l}(\hat{\tau}\mid\eta_{0})+o_p\left(n^{-1/2}\right)$. We apply another Taylor expansion, this time on $\overline{U}_l(\hat{\tau}\mid\eta_{0})$ about $\hat{\tau}=\tau_{0}$, and we obtain
		\begin{equation*}
		\hat{\tau}-\tau_0=-\mathrm{E}\left\{\frac{\partial}{\partial\tau^T}\overline{U_l}(\tau_0\mid\eta_{0})\right\}^{-1}\overline{U_l}(\tau_0\mid\eta_{0})+o_p\left(n^{-1/2}\right).
		\end{equation*}
		Because $\mathrm{E}\left\{S_{obs}(\eta_0;X_\mathrm{obs},A,Y)\right\}=0$, the first term simplifies as
		\begin{align*}
		\mathrm{E}\left\{\frac{\partial}{\partial\tau^T}\overline{U_l}(\tau_0\mid\eta_{0})\right\}&=\mathrm{E}\left\{\frac{\partial}{\partial\tau^T}\overline{U}(\tau_0;X_\mathrm{obs},A,Y,\eta_{0})\right\}\\
		&=\mathrm{E}\left\{\frac{\partial}{\partial\tau^T}\overline{U}(\tau_0;X,A,Y)\right\}\\
		&=\lambda.
		\end{align*}
		
		Lastly, if we combine all the results above, we obtain an asymptotic linearization of $\hat{\tau}-\tau_{0}$ as
		\begin{multline}
		\hat{\tau}-\tau_0=-\lambda^{-1}\{\overline{U_l}(\tau_0|\eta_{0})\nonumber\\
		=-\lambda^{-1} n^{-1}\sum_{i=1}^{n}\left\{\overline{U}(\tau_{0};X_\mathrm{obs,i},A_i,Y_i,\eta_{0})+\kappa S_{obs}(\eta_0;X_\mathrm{obs,i},A_i,Y_i)\right\}+o_p\left(n^{-1/2}\right)
		\end{multline}
		with $\kappa$ and $\lambda$ defined as in the above theorem. 
		
		Therefore, the asymptotic variance of $\sqrt{n}(\hat{\tau}-\tau_{0})$ is
		\begin{equation*}
		\lambda^{-1}\mathrm{Var}\left\{\overline{U}(\tau_{0};X_\mathrm{obs},A,Y,\eta_{0})+\kappa S_{obs}(\eta_0;X_\mathrm{obs},A,Y)\right\}\lambda^{-1},
		\end{equation*}
		which completes the proof.
	\end{proof*}
	
	As a result of Theorem \ref{Thm:ExactVarEst}, we can obtain a consistent variance estimator of $\hat{\tau}$. Define $\hat{\lambda}$ and $\hat{\kappa}$ as empirical versions of $\lambda$ and $\kappa$, respectively. For example, $\hat{\lambda}=n^{-1}\sum_{i=1}^{n}\frac{\partial}{\partial\tau}U(\hat{\tau};X_\mathrm{obs,i},A_i,Y_i,\hat{\eta})$. Next define $\hat{q}_i=\overline{U}(\hat{\tau};X_\mathrm{obs,i},A_i,Y_i,\hat{\eta})+\hat{\kappa} S_{obs}(\hat{\eta};X_\mathrm{obs,i},A_i,Y_i)$. Then we can estimate the variance of $\hat{\tau}$ using a sandwich formula
	\begin{equation*}
	\mathrm{Var}(\hat{\tau})=\hat{\lambda}^{-1}\left\{n^{-1}\frac{1}{n-1}\sum_{i=1}^{n}\left(\hat{q}_i-\overline{\hat{q}}\right)^2\right\}\hat{\lambda}^{-1},
	\end{equation*}
	where $\overline{\hat{q}}=n^{-1}\sum_{i=1}^{n}\hat{q}_i$.
	
	\subsection{Variance Estimation}
	Because of how data is imputed under FI, we can obtain further simplification when using IPW or AIPW. The $\lambda$ terms cancels since $\frac{\partial}{\partial\tau}U(\tau_0;X,A,Y)=-1$ under either estimator, and it can be shown that the $\kappa S_\mathrm{obs}(\eta_{0})$ term falls out of $\varOmega$ for observations where $X_{2}$ is observed.  However, even after these mild simplifications, it is still apparent that exact variance estimates will be difficult to obtain. Only in rare situations will the derivatives of the score functions be anything other than impractical to calculate. Theorem \ref{Thm:ExactVarEst} suggests large sample approximation is appropriate, and a bootstrap or jackknife estimator can serve as a more practical alternative.

	\section{Simulation Study}
	In the current causal inference literature, there has not been a side-by-side comparison of FI and MI with respect to how they perform estimating average treatment effects or of their corresponding variance estimates in the same setting. To examine how FI performs compared to MI, we adapt a simulation setup previously used by \citet{lunceford2004stratification}. We modify this setting for our purposes by creating missingness for covariates. We examine the bias and variance properties of FI compared to two versions of MI as well as CC estimation. Estimation of $\tau_0$ under fully observed data is also conducted as a reference point.
	\subsection{Simulation Setup}
	Let $X=(X_1,X_2,X_3)$ be confounders associated with the treatment effect. We generate $X_3$ from a $Bernoulli(0.2)$, and we generate ($X_1,X_2$) from a bivariate normal $N(\mu_{X_3},\Sigma_{X_3})$ conditional on $X_3$ where
	\begin{center}
		$\mu_1=\begin{pmatrix}
		1 \\ -1
		\end{pmatrix}, \mu_0=\begin{pmatrix}
		-1 \\ 1
		\end{pmatrix},$ and $\Sigma_1=\Sigma_0=\begin{pmatrix}
		1 & 0.5 \\
		0.5 & 1
		\end{pmatrix}.$
	\end{center}
	
	We specify the propensity score as $e(X,\theta)=\{1+e^{(0.3+0.2X_1-0.1X_2-0.1X_3)}\}^{-1}$ and generate the treatment indicator $A_i$ from a Bernoulli$\{e(X_i,\theta)\}$. We choose $\theta$ to ensure the starting propensity scores are well behaved (i.e., between 0.1 and 0.9) and satisfy the sufficient overlap assumption.
	
	We generate the outcome $Y$ as 
	\begin{center}
	$Y=-X_1+X_2-X_3+2A+0.5A\times X_1+0.25A\times X_2+\epsilon_Y$
	\end{center}
	where $\epsilon_Y\sim N(0,1)$.
	The addition of the treatment-covariate interaction terms was implemented to better match the simulation data to what is observed in practice. Note that because $X_1$ and $X_2$ have mean 0, these terms fall out in expectation; however to approximate the true value of $\tau_{0}$, both $Y(1)$ and $Y(0)$ were evaluated for all records so the average difference in potential outcomes could be taken. Simulation results for bias and coverage were calculated for each sample relative to this within-sample $\tau_0$.
	
	To simulate our missingness, we generate $R$ from a Bernoulli($\Phi$) with 
	$$\Phi=P(R=1\mid A,X_1,X_3,Y)=1-\left\{1+e^{(0.25+0.25X_1-0.6X_3+0.5A+0.4Y)}\right\}^{-1}.$$
	The coefficient values in $\Phi$ approximate a missingness rate of 0.33. Table \ref{tab:missingnessFreqs} is a two by two contingency table for the missingness and the treatment assignment, averaging across all simulated data sets.\\
	
	\begin{table}[h!]
	\caption{Average Missingness vs Treatment Assignment}\label{tab:missingnessFreqs}
		\centering
		\begin{tabular}{c|c|c|c|}
			\cline{2-4}
			\multicolumn{1}{l|}{} & \textit{Control} & \textit{Treatment} & \textit{Total} \\ \hline
			\multicolumn{1}{|c|}{Complete Case}  & 0.294  & 0.389 & 0.683 \\ \hline
			\multicolumn{1}{|c|}{Missing $X_2$}  & 0.231  & 0.087 & 0.317 \\ \hline
			\multicolumn{1}{|c|}{Total} & 0.524 & 0.476 & 1.000 \\ \hline
		\end{tabular}
	\end{table}
	\noindent A new $X_2^*$ variable is next constructed according to the response indicator, where
	\begin{center}
		$X_{2}^*=\begin{cases}
		X_{2} &$ if $ R=1 \\
		$missing $&$ else$
		\end{cases}.$\\
	\end{center}
	\noindent The observed data is $Z=(X_1,X_2^*,X_3,A,Y,R)$. 
	
	In FI, to motivate the proposal distribution for the missing $X_2^*$ records, we regress $X_2$ on $X_1$ and $X_3$ based on the complete cases. The proposal distribution $h$ in Step 0 of the FI algorithm is a non-central $t(4)$ distribution with the mean and variance matched with the regression model. In Steps 1 and 2, the outcome model is specified as a linear regression model with predictors $X$, $A$, and their interactions. The propensity score model is the same as the true model $e(X,\theta)$. Under these settings, we generate M=200 imputed values for each missing $X_2^*$. The imputation loop continues until either 250 iterations are performed or all parameters converge within $1\mathrm{x}10^{-6}$. The FI loop produces final FI weights which can then be used to calculate the IPW and AIPW estimates for $\tau_{0}$ under FI. We compare these estimates versus CC and MI estimates. For CC, the process is straightforward. For MI we use the mi package. We examine MI under two settings, one where the outcome is included when imputing covariates (which we have labeled MI1), and the other where it is excluded (MI2). While the literature surrounding MI indicates utilizing the response in imputing covariates is preferred \citep{moons2006using, Sterne2009mi1_vs_mi2, Nguyen2017mi1_vs_mi2}, MI2 has been included for completeness. Lastly, we calculate $\hat{\tau}_{IPW}$ and $\hat{\tau}_{AIPW}$ estimates utilizing the full data (as if no missingness had been introduced) for benchmark comparison. 
	
	This process was run 2000 times. To estimate variances in each iteration for each method, a leave-10-out jackknife was used.
	
	\subsection{Simulation Results}
	Once all simulations had been run for all methods, the results were as follows:
	\begin{table}[ht]
		\centering
		\caption{Mean, Coverage, and Execution Results by Method: IPW Results}
		\begin{tabular}{ccccccc}
			\hline
			\multicolumn{1}{c|}{\textit{Method}} & \multicolumn{1}{c}{\textit{MAD}} & \multicolumn{1}{c}{\textit{MSE}} &
			\multicolumn{1}{c}{\textit{Average JackKnife SE}} & \multicolumn{1}{c}{\textit{Coverage}} \\ \hline
			\multicolumn{1}{c|}{FI} & 0.0644 & 0.0065 & 0.0822 & 95.5\% \\
			\multicolumn{1}{c|}{MI1} & 0.0637 & 0.0065 & 0.0911 & 97.8\% \\
			\multicolumn{1}{c|}{MI2} & 0.0801 & 0.0098 & 0.0912 & 93.4\% \\
			\multicolumn{1}{c|}{CC} & 0.1381 & 0.0246 & 0.0764 & 56.6\%  \\ \hline
			\multicolumn{1}{c|}{\textit{Full}} & \textit{0.0525} & \textit{0.0043} &  \textit{0.0643} & \textit{94.9\%}
		\end{tabular}
	\end{table}
	\noindent Similar results were seen for the AIPW estimates:
	\begin{table}[ht]
		\centering
		\caption{Mean, Coverage, and Execution Results by Method: AIPW Results}
		\begin{tabular}{ccccccc}
			\hline
			\multicolumn{1}{c|}{\textit{Method}} & \multicolumn{1}{c}{\textit{MAD}} & \multicolumn{1}{c}{\textit{MSE}} &
			\multicolumn{1}{c}{\textit{Average JackKnife SE}} & \multicolumn{1}{c}{\textit{Coverage}} \\ \hline
			\multicolumn{1}{c|}{FI} & 0.0644 & 0.0065 & 0.0820 & 95.5\% \\
			\multicolumn{1}{c|}{MI1} & 0.0648 & 0.0065 & 0.0917 & 97.6\% \\
			\multicolumn{1}{c|}{MI2} & 0.0798 & 0.0097 & 0.0922 & 93.5\% \\
			\multicolumn{1}{c|}{CC} & 0.1395 & 0.0250 & 0.0754 & 55.6\%  \\ \hline
			\multicolumn{1}{c|}{\textit{Full}} & \textit{0.0518} & \textit{0.0042} &  \textit{0.0633} & \textit{94.8\%}
		\end{tabular}
	\end{table} 
	
	As expected, both MI and FI do better than CC estimators in all regards. It is worth noting how sizable an impact the exclusion of the response variable $Y$ in the imputation step had on MI. Bias increased by 23\% when not including $Y$. Comparing FI to MI1, FI and MI1 had comparable bias, but FI outperformed MI1 in coverage. The FI coverage is much more in-line with the nominal 95\% coverage used in the confidence interval calculation. Multiple imputation (as viewed as MI1) saw over-coverage compared to the full data results. This can be attributed to an inflated standard error.
	
	As to performance, average execution times for FI, MI1, and MI2 were similar. The average time of execution for each method (to perform both the estimate of $tau_0$ and to complete the jack-knife estimation of its variance) are presented in table \ref{tab:simExecutionTimes}. Execution times for CC and Full were negligible and are not reported.
	
	\begin{table}[!h]
		\centering
        \caption{Average Execution Times by Method} \label{tab:simExecutionTimes}
    \begin{tabular}{c|c}
        \hline
        \textit{Method} & \textit{Execution Time (Mins)} \\ \hline
        FI & 24:46 \\
       MI1 & 23:41 \\
        MI2 & 23:44
    \end{tabular}%
    \end{table}
	
	\section{Application to the National Health and Nutrition Examination Survey Data}
	To illustrate the practical usage, we apply our method to a data set from the U.S. National Health and Nutrition Examination Survey. We estimate the effect of cigarette smoking on blood lead levels with age, gender, race, education, and income used as confounders. Of the confounders, only income was subject to missingness at a rate of 8.5\% overall (6.0\% smokers, 9.2\% non-smokers). In the published data set, missing income values were imputed using mean imputation which risks being biased under MAR \citep{NRC2010missingness}. We investigate how the MI and FI estimates change from the analysis based on mean imputation. CC was also examined for a common point of reference.
	
	As in our simulation, we will need to model our propensity scores, as well as regress both our Y (lead level) on all confounders as well as our missing confounder (income) on all present confounders.
	With respect to our regression for income in FI, we built a model of the form
	\begin{center}
	    $X_{inc}=\alpha_0+\alpha_1X_{age}+\alpha_2X_{male}+\alpha_{edu}^TX_{edu}+\alpha_{race}^TX_{race}+\epsilon$
	\end{center}
	where $\alpha_{edu}$ and $X_{edu}$ represent the parameter estimates and data for the dummy variables that represent the 6 education levels (similarly for $\alpha_{race}$ and $X_{race}$ for its 5 levels of race) and $\epsilon$ is an error term with mean $0$. The dummy variables for unknown education and other race were excluded as they were, by construction, linearly dependent on the other columns in their group. This regression also gave us an estimate of $\sigma_{X_{inc}}^{(init)}$. As before, this was used to create M=200 imputed data values for each of the 285 missing cases drawn as $X_{inc;ij}^*=X_{inc;i}^{(\mu)}+\tilde{t}\sigma_{X_{inc}}^{(init)}$ where $\tilde{t}$ was drawn from a t(4) distribution. $h_{X_{inc}}$ was calculated similarly as before. A leave-10-out jackknife was still used to estimate variances.
	
	For MI, the same regression as used in FI was performed but with the response and treatement added in as linear terms. The regression equation for income used under MI was then:
	\begin{center}
	    $X_{inc}=\alpha_0+\alpha_1X_{age}+\alpha_2X_{male}+\alpha_{edu}X_{edu}+\alpha_{race}X_{race}+\alpha_{smoke}A+\alpha_{lead}Y+\epsilon$
	\end{center}
	where $A$ is the treatment indicator for smoking and $Y$ the reported lead level in the blood and $\epsilon$ is an error term with mean $0$. Initial values for the mean and variance of $X_{inc}$ could then be constructed using the complete cases. Imputed values of income were then drawn from the predicted posterior distribution and then subsequently updated via an MCMC process.
	
	The resulting IPW and AIPW estimates and the accompanying variances for $\tau$ can be found in table \ref{tab:SmokeResults}. Additionally, the execution time for each method is provided.
	
	\begin{table}[!ht]
		\centering
		\caption{Results for estimating the effect of cigarette smoking on blood lead levels: estimate (Est), standard error (SE), and execution time}
		\label{tab:SmokeResults}
		\begin{tabular}{l|cc|cc|l}
			& \multicolumn{2}{c|}{IPW} & \multicolumn{2}{c|}{AIPW} &                \\\hline
			\textit{Method} & \textit{Est} & SE &\textit{Est} & \textit{SE} & \textit{Execution Time (Hrs)} \\ \hline
			FI       & 1.290      & 0.231      & 0.932      & 0.163     & 0:59:05   \\
			MI1      & 1.281    & 0.257      & 0.929      & 0.172     & 1:30:18    \\
			CC       &  1.163     & 0.228      & 0.942 &  0.184    & 0:00:18               \\
			\hline
			Original &  1.256     &   0.210    & 0.924 &  0.155         &  N/A         
		\end{tabular}
	\end{table}
	
	While we can not know the true values of income from which we could calculate our bias, an examination of resource utilization does prove useful. It is expected that CC would take minimal time since it does not need to attempt any convergence loop. FI and MI take significantly longer, but we would argue the increase in time is worth the added asymptotic bias advantages over CC. In comparing execution times, MI took about 50\% longer to finish the full mean and jack-knife execution process than FI. As to why this resource gain exists for FI over MI, we postulate that since FI does not have to model and redraw from the full distribution of $X$ every iteration, it is able to arrive at estimates of $\tau_{0}$ and standard errors much more quickly than MI.
	
	\section{Summary and Future Work}
	We have demonstrated that FI is an effective method for addressing missingness in covariates when estimating average treatment effects under the condition that covariates are MAR via simulation study and real-world application. Moreover, we were able to demonstrate FI's superiority over the existing leading methods of MI and CC. FI produces lower bias and better coverage properties than either MI or CC. We also showed that when deployed in a real-world setting, FI is less resource-intensive than MI-based methods, most likely due to the lack of need to estimate the full covariate distribution.
	
	With these results in mind, it is worth noting a comment made by \citet{rubin1996multiple} in his 18 year retrospective of his original work introducing MI. As an initial defense to then contemporary critiques of the method, some of which have been cited here already (see \citet{fay1992inferences} and \citet{meng1994multiple}), Rubin offered up the response that in cases where randomization validity (i.e., actual confidence coverage = nominal interval coverage) is difficult to achieve, statisticians should alternatively seek confidence validity (i.e., actual interval coverage $\ge$ nominal interval coverage) with decisions between competing methods decided by which method has the shortest interval. Near the end of that defense the reader can find the following comment:
	\begin{quotation}
		\textit{Of course, if we have a procedure that is confidence valid but not randomization valid, there is hope that a better confidence-valid procedure exists (i.e., one with shorter intervals), which is also randomization valid, but in general this is not achievable \citep[475]{rubin1996multiple}.}
	\end{quotation}

	It is our belief the results above demonstrate FI produces randomization valid inference based on general estimation approaches for the population average treatment effect that MI may lack.
	
	In our own future work, we will extend the FI algorithm in Section 3.1 to the MNAR setting by the inclusion of a model for the missingness and considering the full likelihood from equation (\ref{eq:full_decomp}). When the covariates are MNAR, an important challenge is that the full likelihood function (\ref{eq:full_decomp}) is not identifiable (or recoverable under \citet{mohan2018graphical}) in general. To overcome this challenge, one may consider non-response instrumental variable methods (e.g., \citet{yang2019causal}), missingness graphical models (e.g., \citet{mohan2018graphical}), negative controls (e.g., \citet{KurokiPearl2014negControl}), and sensitivity analysis (e.g., \citet{cornfield1959Smoking}). Once the identification conditions are established, the FI algorithm can be developed similarly based on the full likelihood equation (\ref{eq:full_decomp}). Finally, we would like to explore FI's potential uses in other causal inference methods for treatment effect estimation beyond weighting methods, particularly as it applies to matching methods.
	\bibliographystyle{dcu}
	\bibliography{JCI_FIvMI.bib}
	\newpage
	
	\global\long\def\theequation{A\arabic{equation}}
 \setcounter{equation}{0}

\global\long\def\thelemma{A\arabic{lemma}}
 \setcounter{lemma}{0}

\global\long\def\theexample{A\arabic{example}}
 \setcounter{example}{0}

\global\long\def\thesection{A\arabic{section}}
 \setcounter{section}{0}

\global\long\def\thetheorem{A\arabic{theorem}}
 \setcounter{theorem}{0}

\global\long\def\theremark{A\arabic{remark}}
 \setcounter{remark}{0}

\global\long\def\thestep{A\arabic{step}}
 \setcounter{step}{0}

\global\long\def\theassumption{A\arabic{assumption}}
 \setcounter{assumption}{0}

\global\long\def\theproof{A\arabic{proof}}

	\section{Appendix}
	\subsection{Identifiability of Average Treatment Effect under MAR}
	
	\paragraph*{Proof of Theorem 1}
	It is well known that under MAR, the full data distribution is identifiable from the observed data. 
	To show that the average treatment effect $\tau_0$ can be identified from the observed data, we express
	\begin{eqnarray*}
		\tau_0 & = & E\left\{ Y(1)-Y(0)\right\} \\
		& = & E\left\{ \frac{AY}{e(X)}-\frac{(1-A)Y}{1-e(X)}\right\} \\
		%& = & E\bigg[\int_{X_{\mathrm{mis}},\mathcal{R}}\left\{ \frac{AY}{e(X_{\mathrm{obs}},x_{\mathrm{mis}})}-\frac{(1-A)Y}{1-e(X_{\mathrm{obs}},x_{\mathrm{mis}})}\r%ight\}\\
		%& & \hspace{1.5cm} f(X_{mis}=x_{\mathrm{mis}},R=r\mid %X_{\mathrm{obs},}A,Y)\mathrm{d}x_{\mathrm{mis}}\mathr%m{d}r\bigg]\\
		& = & E\bigg[\int_{{X}_{{mis}}}\left\{ \frac{AY}{e(X_{\mathrm{obs}},x_{\mathrm{mis}})}-\frac{(1-A)Y}{1-e(X_{\mathrm{obs}},x_{\mathrm{mis}})}\right\} f(X_{\mathrm{mis}}=x_{\mathrm{mis}}\mid X_{\mathrm{obs},}A,Y)\mathrm{d}x_{\mathrm{mis}}\bigg],
	\end{eqnarray*}
	where the second equality follows by Assumption 1 and 2, and the fourth
	equality follows by Assumption 3. 
	
	\subsection{Example FI implementatoin}
	\begin{example}
	[Generating a complete data set under FI]\label{example:Toy_FI}
	    For illustration, consider a population where covariates $X_1$ and $X_2$, treatment indicator $A$, and response indicator $Y$ are all binomial variables. Let $X_1\sim Binomial(0.5)$. Let $X_2\sim Binomial(\phi_{X_1})$ where $\phi_{X_1}=0.4+0.2X_1$. Let $A\sim Binomial\left\{(1+e^{X_1+X_2})^{-1}\right\}$. Finally, let $Y\sim Binomial\left\{(1+e^{X_1+X_2-2A})^{-1}\right\}$.
	    
	   Suppose we draw a sample of size $n=10$ from the population described above but some observation are missing values for $X_2$ with probability $pR=(1+e^{X_1+A+Y})^{-1}$, leaving us with the starting data set in Figure \ref{tab:Toy-StartData}.
        \begin{table}[h!]
        \centering
        \caption{Pre-imputed data. Observation 2 and observation 3 are both missing values for $X_2$.}
        \label{tab:Toy-StartData}
        \begin{tabular}{l|cccc}
        \textbf{ID} & \textbf{$X_1$} & \textbf{$X_2$} & \textbf{$A$} & \textbf{$Y$} \\ \hline
        1 & 1 & 0 & 0 & 0 \\
        2 & 1 & N/A & 0 & 0 \\
        3 & 0 & N/A & 0 & 0 \\
        4 & 0 & 1 & 0 & 0 \\
        5 & 0 & 1 & 0 & 1 \\
        6 & 0 & 1 & 0 & 0 \\
        7 & 1 & 0 & 1 & 0 \\
        8 & 0 & 0 & 1 & 1 \\
        9 & 1 & 1 & 0 & 0 \\
        10 & 0 & 1 & 0 & 0
        \end{tabular}
        \end{table}
        
        To generate imputed values for units 2 and 3, a proposal distribution must be provided. Adopting the decomposition from equation \ref{eq:MAR_decomp}, a natural choice is $h(X_2)=f(X_2\mid X_1;\alpha)$. Since $X_2$ is binomial, we can estimate $f(X_2\mid X_1;\alpha)$ via logistic regression. At this point, $\alpha$ can only be estimated based on the complete cases. This results in the proposal distribution
        
        \[
        P(X_2=1)=\frac{1}{1+e^{-(1.386-2.079X_1)}},
        \]

        \noindent with $\hat{\alpha}=(\hat{\alpha}_0,\hat{\alpha}_1)=(1.386,-2.079)$ or put more plainly the probability mass function corresponding to Table \ref{tab:Toy-h0_pmf}.
         \begin{table}[h!]
        \centering
        \caption{PMF for imputation function h().}
        \label{tab:Toy-h0_pmf}
        \begin{tabular}{ccc}
        \textbf{$X_1$} & \textbf{$X_2$} & \textbf{$f(X_2\mid X_1;\hat{\alpha})$} \\ \hline
        1 & 1 & 0.3333  \\
        1 & 0 & 0.6667  \\
        0 & 1 & 0.8000  \\
        0 & 0 & 0.2000  
        \end{tabular}
        \end{table}
        
        With $h()$ in hand, we can generate $M$ new values for each missing $X_2$. If for the sake of space, we let $M=5$, then this would mean we would generate $5$ values $(X_{2,2}^{*(j)}; j=1,...,5)$ for observation 2 drawing from $Binomial(0.3333)$ and $5$ values $(X_{2,3}^{*(j)}; j=1,...,5)$ for observation 3 drawing from $Binomial(0.8000)$. At this point in the process, no weights have been updated, so the fractional weight on each of these imputed values is $1/M$, leaving us with our initial imputed data set (Table \ref{tab:Toy-dat0}) to conclude the I-step. Note, since the imputed values $X_2^*$ never change, the values $h(X_2^*)$ never change, so it is convenient to include a column holding these values, as they will be used repeatedly when updating the fractional weights in each W-step. In Table \ref{tab:Toy-dat0} the column holding the proposal distribution likelihood for each observation is labeled $h_0$.
        
        \begin{table}[h!]
        \centering
        \caption{Initial imputed data set prior to the first W-step. }
        \label{tab:Toy-dat0}
        \begin{tabular}{l|llllll}
        ID & \textbf{$X_1$} & \textbf{$X_2^{*}$} & \textbf{A} & \textbf{Y} & \textbf{$\omega^{*}$} & \textbf{$h_0$} \\ \hline
        1 & 1 & 0 & 0 & 0 & 1 & 0.6667 \\
        2.1 & 1 & 0 & 0 & 0 & 0.2 & 0.6667 \\
        2.2 & 1 & 1 & 0 & 0 & 0.2 & 0.3333 \\
        2.3 & 1 & 0 & 0 & 0 & 0.2 & 0.6667 \\
        2.4 & 1 & 1 & 0 & 0 & 0.2 & 0.3333 \\
        2.5 & 1 & 1 & 0 & 0 & 0.2 & 0.3333 \\
        3.1 & 0 & 0 & 0 & 0 & 0.2 & 0.2000 \\
        3.2 & 0 & 0 & 0 & 0 & 0.2 & 0.2000 \\
        3.3 & 0 & 0 & 0 & 0 & 0.2 & 0.2000 \\
        3.4 & 0 & 1 & 0 & 0 & 0.2 & 0.8000 \\
        3.5 & 0 & 1 & 0 & 0 & 0.2 & 0.8000 \\
        4 & 0 & 1 & 0 & 0 & 1 & 0.8000 \\
        5 & 0 & 1 & 0 & 1 & 1 & 0.8000 \\
        6 & 0 & 1 & 0 & 0 & 1 & 0.8000 \\
        7 & 1 & 0 & 1 & 0 & 1 & 0.6667 \\
        8 & 0 & 0 & 1 & 1 & 1 & 0.2000 \\
        9 & 1 & 1 & 0 & 0 & 1 & 0.3333 \\
        10 & 0 & 1 & 0 & 0 & 1 & 0.8000
        \end{tabular}
        \end{table}
        
        To begin the first W-step, we must first estimate the full joint distribution of $(X_1,X_2,A,Y)$. Again we will return to the decomposition in equation \ref{eq:MAR_decomp}, and because both $A$ and $Y$ are binomial we will choose to estimate the distributions $e(X_1,X_2\mid\theta)$ and $f(Y\mid X_1,X_2,A;\beta)$ using weighted logistic regression. Similarly, from now on we will be estimating $f(X_2\mid X_1;\alpha)$ using weighted logistic regression to incorporate the fractional weights (whereas before, only the complete cases were used in the I-step). Initial values of $\hat{\alpha}^{(t)}$, $\hat{\beta}^{(t)}$, $\hat{\theta}^{(t)}$ for $t=1$ are as follows:
        
        \begin{itemize}
            \item $\hat{\alpha}^{(1)}=(\hat{\alpha}_0,\hat{\alpha}_1)=(1.0116,-1.4171)$,
            \item $\hat{\beta}^{(1)}=(\hat{\beta}_0,\hat{\beta}_1,\hat{\beta}_2,\hat{\beta}_A)=(-19.7160,-39.1894,18.4922,39.3107)$,
            \item $\hat{\theta}^{(1)}=(\hat{\theta}_0,\hat{\theta}_1,\hat{\theta}_2)=(0.5108,-0.8473,-20.8207)$.
        \end{itemize}
        
        Once initial values for $(\hat{\alpha},\hat{\beta},\hat{\theta})$ are estimated, it is convenient to calculate the likelihood of each observation's $X_2^*$, $A$, and $Y$ under their respective current estimated distributions, that is to say calculating $P(X_2=X_{2,i}^{*(j)}\mid\hat{\alpha}^{(1)})$ for all observations and similarly for $A$ and $Y$ with their distributions and current parameters. For short hand, let $f_{X_2}()$, $f_A()$, and $f_Y()$ be the likelihood functions for $X_2$, $A$, and $Y$ respectively under the current parameter values $\hat{\alpha}^{(t)}$, $\hat{\beta}^{(t)}$, $\hat{\theta}^{(t)}$ that were just calculated. For example, $f_Y(Y_i)=f(Y_i\mid A_i, X_{1,i}, X _{2,i} ^{*(j)};\hat{\beta}^{(t)}).$ With the respective likelihoods calculated, each observation's fractional weight can be updated as
       
        \[
        \omega_i^{*(j)}=\frac{f_{X_2}(X_{2,i}^{*(j)})f_A(A_i)f_Y(Y_i)}{h_{0,ij}}.
        \]

        It is possible at this point that $\sum_{j=1}^{M}\omega_i^{*(j)}\neq 1$ for an individual. If so the weights for that individual will need re-normalized to ensure the condition $\sum_{j=1}^{M}\omega_i^{*(j)}=1$ is still met $\forall i$. The updated data set after the first W-step can be seen in Table \ref{tab:Toy-dat1}. Note the only data elements to change between Table \ref{tab:Toy-dat0} and Table \ref{tab:Toy-dat1} are the fractional weights in column $\omega^*$.
        
        \begin{table}[h!]
        \centering
        \caption{Updated data after the first W-step.}
        \label{tab:Toy-dat1}
        \begin{tabular}{l|llllll}
        ID & \textbf{$X_1$} & \textbf{$X_2^{*}$} & \textbf{A} & \textbf{Y} & \textbf{$\omega^{*}$} & \textbf{$h_0$} \\ \hline
        1 & 1 & 0 & 0 & 0 & 1 & 0.6667 \\
        2.1 & 1 & 0 & 0 & 0 & 0.1129 & 0.6667 \\
        2.2 & 1 & 1 & 0 & 0 & 0.2581 & 0.3333 \\
        2.3 & 1 & 0 & 0 & 0 & 0.1129 & 0.6667 \\
        2.4 & 1 & 1 & 0 & 0 & 0.2581 & 0.3333 \\
        2.5 & 1 & 1 & 0 & 0 & 0.2581 & 0.3333 \\
        3.1 & 0 & 0 & 0 & 0 & 0.1714 & 0.2000 \\
        3.2 & 0 & 0 & 0 & 0 & 0.1714 & 0.2000 \\
        3.3 & 0 & 0 & 0 & 0 & 0.1714 & 0.2000 \\
        3.4 & 0 & 1 & 0 & 0 & 0.2429 & 0.8000 \\
        3.5 & 0 & 1 & 0 & 0 & 0.2429 & 0.8000 \\
        4 & 0 & 1 & 0 & 0 & 1 & 0.8000 \\
        5 & 0 & 1 & 0 & 1 & 1 & 0.8000 \\
        6 & 0 & 1 & 0 & 0 & 1 & 0.8000 \\
        7 & 1 & 0 & 1 & 0 & 1 & 0.6667 \\
        8 & 0 & 0 & 1 & 1 & 1 & 0.2000 \\
        9 & 1 & 1 & 0 & 0 & 1 & 0.3333 \\
        10 & 0 & 1 & 0 & 0 & 1 & 0.8000
        \end{tabular}
        \end{table}
        
        blueThe M-step, follows naturally from the W-step. Given the updated weights $\omega^*$ in Table \ref{tab:Toy-dat1}, parameters $\hat{\alpha}^{(t)}$, $\hat{\beta}^{(t)}$, and $\hat{\theta}^{(t)}$ are all updated from $t=1$ to $t=2$ again using weighted logistic regression. In the first M-step that would mean updating $\hat{\alpha}^{(1)}$, $\hat{\beta}^{(1)}$, $\hat{\theta}^{(1)}$ to
        
        \begin{itemize}
            \item $\hat{\alpha}^{(2)}=(\hat{\alpha}_0,\hat{\alpha}_1)=(1.0860,-1.3127)$,
            \item $\hat{\beta}^{(2)}=(\hat{\beta}_0,\hat{\beta}_1,\hat{\beta}_2,\hat{\beta}_A)=(-19.7179,-39.2118,18.4692,39.3238)$,
            \item $\hat{\theta}^{(2)}=(\hat{\theta}_0,\hat{\theta}_1,\hat{\theta}_2)=(0.6650,-0.8686,-20.9515)$.
        \end{itemize}
        
        The updated parameters are checked against some convergence criteria (for instance stopping once the max absolute difference between all parameters is $\leq 0.0001$). If the convergence criteria is not met, the process cycles back to the W-step, and weights are recalculated now using parameters $\hat{\alpha}^{(2)}$, $\hat{\beta}^{(2)}$, $\hat{\theta}^{(2)}$. With new weights the M-step will update distributional parameters $\hat{\alpha}^{(t)}$, $\hat{\beta}^{(t)}$, $\hat{\theta}^{(t)}$, and the process continues until the convergence criteria are met. In our sample, the parameters converged on the $18^{th}$ iteration, producing the final FI-complted data set in Table \ref{tab:Toy-dat_final}. This data set can now be incorporated into any weighted analysis where a complete-data consistent estimator exists
        
        \begin{table}[h!]
        \centering
        \caption{Final FI-completed data set.}
        \label{tab:Toy-dat_final}
        \begin{tabular}{l|lllll}
        ID & \textbf{$X_1$} & \textbf{$X_2^{*}$} & \textbf{A} & \textbf{Y} & \textbf{$\omega^{*}$} \\ \hline
        1 & 1 & 0 & 0 & 0 & 1 \\
        2.1 & 1 & 0 & 0 & 0 & 0.0886 \\
        2.2 & 1 & 1 & 0 & 0 & 0.2743 \\
        2.3 & 1 & 0 & 0 & 0 & 0.0886 \\
        2.4 & 1 & 1 & 0 & 0 & 0.2743 \\
        2.5 & 1 & 1 & 0 & 0 & 0.2743 \\
        3.1 & 0 & 0 & 0 & 0 & 0.1334 \\
        3.2 & 0 & 0 & 0 & 0 & 0.1334 \\
        3.3 & 0 & 0 & 0 & 0 & 0.1334 \\
        3.4 & 0 & 1 & 0 & 0 & 0.3000 \\
        3.5 & 0 & 1 & 0 & 0 & 0.3000 \\
        4 & 0 & 1 & 0 & 0 & 1 \\
        5 & 0 & 1 & 0 & 1 & 1 \\
        6 & 0 & 1 & 0 & 0 & 1 \\
        7 & 1 & 0 & 1 & 0 & 1 \\
        8 & 0 & 0 & 1 & 1 & 1 \\
        9 & 1 & 1 & 0 & 0 & 1 \\
        10 & 0 & 1 & 0 & 0 & 1
        \end{tabular}
        \end{table}
        
	\end{example}

	\subsection{Sensitivity to Imputation Size}
	In our primary simulation study, the imputation size $M$ was established to be 200 with the assumption that such an $M$ would be sufficiently large to obtain approximately asymptotic results. Traditional recommendations for the size of $M$ are much smaller ($M=2$ to $M=10$) when only inference about the point estimates were of interest. More recent recommendations for the selection of $M$ have occurred when also needing accurately estimate variance of the point estimate are somewhat higher. For our data set with \%-missingness approximately equal to 0.33, recommendations for an $M$ size in the context of MI range from 20-40 \citep{Graham2007_MI_Msize,vonHippel2016_MI_Msize}. To examine if we could lower our $M$ setting for FI we reran the first 500 simulations of our analysis varying $M$ to be $M=5,10,20,50,100$. The results are summarized below in both figure 1 and in table 5.

	\begin{figure}[ht]
		\caption{Comparing sensitivity to size of M on bias and coverage among FI and MI implementations when estimating treatment effects via IPW}
		\includegraphics[width=.48\linewidth]{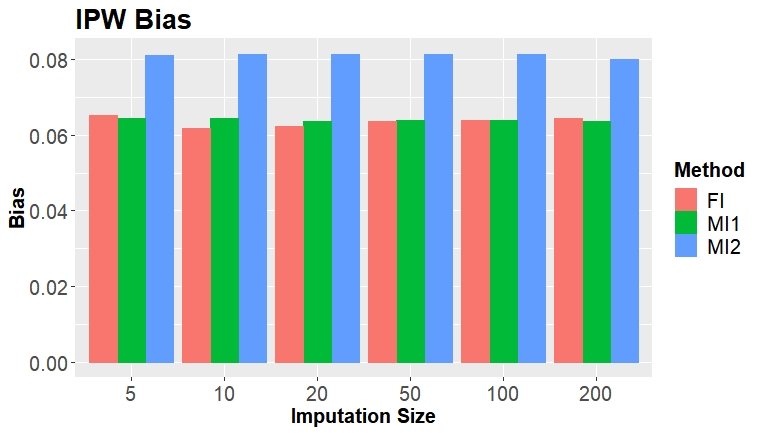}
		\includegraphics[width=.48\linewidth]{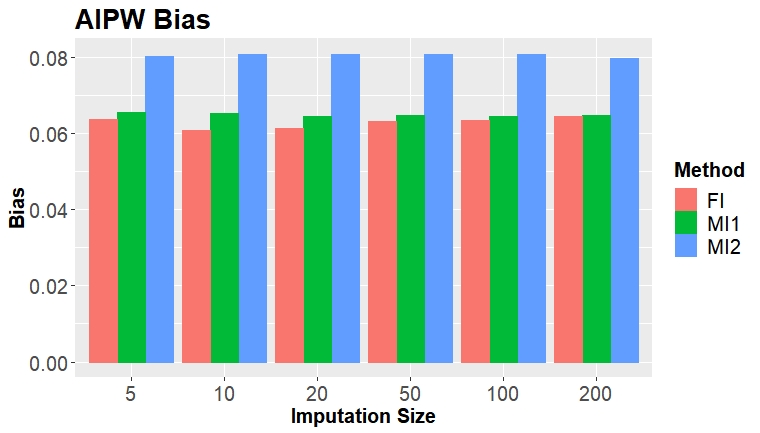} \\
		\includegraphics[width=.48\linewidth]{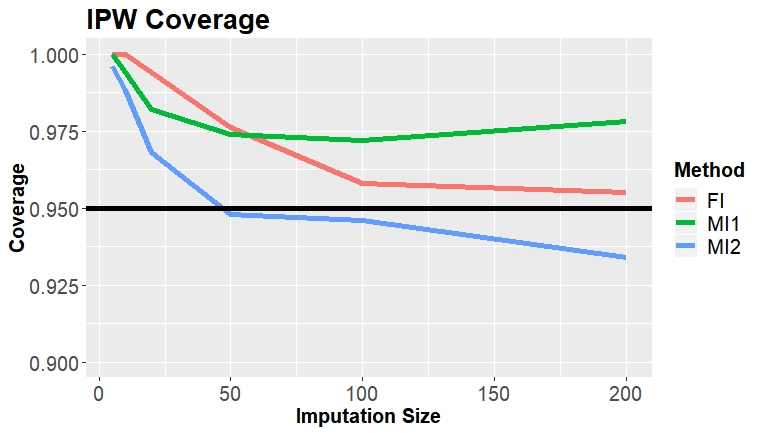}
		\includegraphics[width=.48\linewidth]{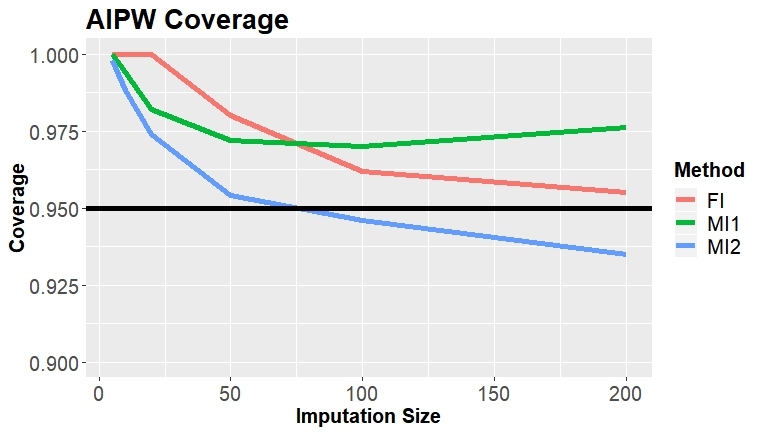}
	\end{figure}

	\begin{table}[ht]
		\centering
		\caption{Table of results for comparing sensitivity to size of M on bias and coverage among FI and MI implementations}
		\begin{tabular}{c|c|c|c|c|}
			\cline{2-5}
			& \multicolumn{2}{c|}{\textit{IPW}} & \multicolumn{2}{c|}{\textit{AIPW}} \\ \hline
			\multicolumn{1}{|c|}{\textit{Setting}} & \textit{Bias} & \textit{Coverage} & \textit{Bias} & \textit{Coverage} \\ \hline
			\multicolumn{1}{|c|}{FI\_M5} & 0.065 & 100.0\% & 0.064 & 100.0\% \\ \hline
			\multicolumn{1}{|c|}{MI1\_M5} & 0.065 & 100.0\% & 0.065 & 100.0\% \\ \hline
			\multicolumn{1}{|c|}{MI2\_M5} & 0.081 & 99.6\% & 0.080 & 99.8\% \\ \hline \hline
			\multicolumn{1}{|c|}{FI\_M10} & 0.062 & 100.0\% & 0.061 & 100.0\% \\ \hline
			\multicolumn{1}{|c|}{MI1\_M10} & 0.064 & 99.4\% & 0.065 & 99.4\% \\ \hline
			\multicolumn{1}{|c|}{MI2\_M10} & 0.082 & 98.8\% & 0.081 & 98.8\% \\ \hline \hline
			\multicolumn{1}{|c|}{FI\_M20} & 0.062 & 99.4\% & 0.061 & 100.0\% \\ \hline
			\multicolumn{1}{|c|}{MI1\_M20} & 0.064 & 98.2\% & 0.065 & 98.2\% \\ \hline
			\multicolumn{1}{|c|}{MI2\_M20} & 0.081 & 96.8\% & 0.081 & 97.4\% \\ \hline \hline
			\multicolumn{1}{|c|}{FI\_M50} & 0.064 & 97.6\% & 0.063 & 98.0\% \\ \hline
			\multicolumn{1}{|c|}{MI1\_M50} & 0.064 & 97.4\% & 0.065 & 97.2\% \\ \hline
			\multicolumn{1}{|c|}{MI2\_M50} & 0.082 & 94.8\% & 0.081 & 95.4\% \\ \hline \hline
			\multicolumn{1}{|c|}{FI\_M100} & 0.064 & 95.8\% & 0.063 & 96.2\% \\ \hline
			\multicolumn{1}{|c|}{MI1\_M100} & 0.064 & 97.2\% & 0.065 & 97.0\% \\ \hline
			\multicolumn{1}{|c|}{MI2\_M100} & 0.082 & 94.6\% & 0.081 & 94.6\% \\ \hline \hline
			\multicolumn{1}{|c|}{FI\_M200*} & 0.064 & 95.5\% & 0.064 & 95.5\% \\ \hline
			\multicolumn{1}{|c|}{MI1\_M200*} & 0.064 & 97.8\% & 0.065 & 97.6\% \\ \hline
			\multicolumn{1}{|c|}{MI2\_M200*} & 0.080 & 93.4\% & 0.080 & 93.5\% \\ \hline
		\end{tabular}%
	\end{table}
	
	Our results demonstrate no loss of accuracy between FI and MI1 regardless of the setting of $M$. There also is no gain in accuracy among any of the methods with increasing $M$. This matches with existing literature that if only point estimation is of interest then low $M$ settings are sufficient for MI. Our results further suggest FI can also permit low $M$ settings when variance estimation is not of interest. However, all methods perform poorly with respect to coverage until $M$ is at least greater than 20. As \cite{Graham2007_MI_Msize} and \cite{vonHippel2016_MI_Msize} suggest, low $M$ values are not sufficient for accurate variance estimation. Moreover, if focusing on MI1, the coverage gains stop somewhere between $M=20$ and $M=50$ (again, as expected), but the MI1 coverage is still about 2\% higher than the nominal coverage.
	
	As for FI, it takes slightly higher $M$ to surpass the coverage potential of MI1. At $M=100$, FI coverage was within 1\% of nominal coverage with gains still being seen as $M$ increased to the settings used in our simulations. From these results, we conclude $M=200$ was a sufficient setting for our simulations though even better coverage may have been attainable with higher $M$. Furthermore, if computational resources are limited setting $M$ to only 100 may be a passable setting -- still superior to MI but not yet reaching approximate asymptotic properties. It is also important to note that these sensitivity results are only confirmatory for our simulation settings and may differ when FI and MI are applied in more complex settings or when there is a higher level of missingness. As such, we recommend plotting coverage curves like these when deploying either method in future applications to validate $M$ has been set sufficiently high in those situations.

\end{document}